\documentclass[pra,twocolumn,superscriptaddress]{revtex4-1}

\usepackage{graphicx}
\usepackage{amsmath}
\usepackage{amssymb}
\usepackage{hyperref}
\usepackage{braket}
\usepackage{ulem}
\usepackage{bbm}
\usepackage[utf8]{inputenc}
\usepackage[T1]{fontenc}
\usepackage{xcolor}

\begin{document}

\title{Long distance spin shuttling enabled by few-parameter velocity optimization}
\author{Alessandro David}
\affiliation{Peter Grünberg Institute-Quantum Control (PGI-8), Forschungszentrum Jülich GmbH, D-52425 Jülich, Germany}
\author{Akshay Menon Pazhedath} 
\affiliation{Peter Grünberg Institute-Quantum Control (PGI-8), Forschungszentrum Jülich GmbH, D-52425 Jülich, Germany}
\affiliation{Institute for Theoretical Physics, University of Cologne, Zülpicher Straße 77, 50937 Cologne, Germany}
\author{Lars R. Schreiber} 
\affiliation{JARA-FIT Institute for Quantum Information, Forschungszentrum Jülich GmbH and RWTH Aachen University, Aachen, Germany}
\affiliation{ARQUE Systems GmbH, 52074 Aachen, Germany}
\author{Tommaso Calarco}
\affiliation{Peter Grünberg Institute-Quantum Control (PGI-8), Forschungszentrum Jülich GmbH, D-52425 Jülich, Germany}
\affiliation{Institute for Theoretical Physics, University of Cologne, Zülpicher Straße 77, 50937 Cologne, Germany}
\affiliation{Dipartimento di Fisica e Astronomia, Università di Bologna, 40127 Bologna, Italy}
\author{Hendrik Bluhm} 
\affiliation{JARA-FIT Institute for Quantum Information, Forschungszentrum Jülich GmbH and RWTH Aachen University, Aachen, Germany}
\affiliation{ARQUE Systems GmbH, 52074 Aachen, Germany}
\author{Felix Motzoi}
\affiliation{Peter Grünberg Institute-Quantum Control (PGI-8), Forschungszentrum Jülich GmbH, D-52425 Jülich, Germany}
\affiliation{Institute for Theoretical Physics, University of Cologne, Zülpicher Straße 77, 50937 Cologne, Germany}

\begin{abstract}
Spin qubit shuttling via moving conveyor-mode quantum dots in Si/SiGe offers a promising route to scalable miniaturized quantum computing.
Recent modeling of dephasing via valley degrees of freedom and well disorder dictate a slow shutting speed which seems to limit errors to above correction thresholds if not mitigated.
We increase the precision of this prediction, showing that typical errors for 10~\textmu m shuttling at constant speed results in O(1) error, using fast, automatically differentiable numerics and including improved disorder modeling and potential noise ranges. However, remarkably, we show that these errors can be brought to well below fault-tolerant thresholds using trajectory shaping with very simple parametrization with as few as 4 Fourier components, well within the means for experimental in-situ realization, and without the need for targeting or knowing the location of valley near degeneracies.
\end{abstract}

\maketitle

\section{Introduction}

Scalability is one of the fundamental requirements for a functional quantum computer \cite{divincenzo_physical_2000} and in semiconductor based platforms \cite{burkard_semiconductor_2023} the quest for a fault-tolerant scalable architecture is at a crucial point \cite{vandersypen_interfacing_2017}. Spin qubits in quantum dots have recently achieved single- and two-qubit gate fidelities above the error correction threshold \cite{yoneda_quantum-dot_2018, noiri_fast_2022, xue_quantum_2022, mills_two-qubit_2022}. Moreover, they are compatible with industrial fabrication processes inheriting large-scale production and characterization techniques \cite{neyens_probing_2024, huckemann_prep_2024}. The small size factor of these qubits theoretically allows to host millions of them on a single chip, but the architecture's short-ranged exchange interaction needs to be supplied with a coherent channel to move the qubits around and to host on-chip electronics \cite{vandersypen_interfacing_2017}.

Long distance qubit transfer through gate defined quantum dot arrays has been recently well characterized \cite{langrock_blueprint_2023}. We focus here on the conveyor-mode regime in Si/SiGe heterostructures which has shown promising results for charge transport \cite{seidler_conveyor-mode_2022,xue_sisige_2024} as well as spin transport fidelities for short distances \cite{struck_spin-epr-pair_2024,de_smet_high-fidelity_2024}. One of the key challenges faced by this transfer approach is the presence of a valley degree of freedom in silicon \cite{zwanenburg_silicon_2013} which interacts with the moving spin through valley-dependent $g$-factors \cite{kawakami_electrical_2014, veldhorst_spin-orbit_2015, ferdous_valley_2018, ruskov_electron_2018}. A new model for the intervalley coupling \cite{paquelet_wuetz_atomic_2022, lima_interface_2023, losert_practical_2023, lima_valley_2024} has highlighted how interface disorder determines rapid changes in the valley mixing leading to excited valley population and, therefore, spin dephasing with high probability during shuttling unless the spots with very low valley splitting are carefully avoided \cite{losert_strategies_2024}. In addition, the valley relaxes to the groundstate \cite{yang_spin-valley_2013, tahan_relaxation_2014} with measured lifetimes above 10 ms \cite{penthorn_direct_2020} for low valley splitting values.

In this work we simulate conveyor-mode spin shuttling by building a comprehensive picture over several experimental parameters and we show how few harmonic corrections to the trajectory lead to sub-fault-tolerant shuttling fidelities compared to the low fidelity values of plain constant speed shuttling. We obtain realistic numerical results by generating position-dependent valley Hamiltonians derived from the alloy diffusion model \cite{paquelet_wuetz_atomic_2022}. Spin and valley are coupled through the $g$-factor valley-dependence and their joint state is propagated following open system dynamics, exploring a variety of relevant parameter regimes.

We proceed to optimize the quantum dot shuttling speed as a function of time with an automatically-differentiated optimization method on a sinusoidal basis in order to decrease the spin shuttling infidelity. The method allows maintaining high spin purity at the end of the shuttling by removing the entanglement with the valley. We find that for expected valley relaxation times, namely in the order of 100~$\mu$s, a shuttling infidelity around $10^{-3}$ is reachable with as few as 4 frequency components.

This paper is structured as follows. In Sec.~\ref{sec:model} we present our model, the open system dynamics and the details of the position-dependent alloy model. In Sec.~\ref{sec:optimization} we describe how we perform the optimization and in Sec.~\ref{sec:results} we report our results for constant speed and optimized shuttling. Finally, in Sec.~\ref{sec:Conclusions} we draw our conclusions.

\section{Model}\label{sec:model}

\begin{figure}
    \centering
    \includegraphics[width=\columnwidth]{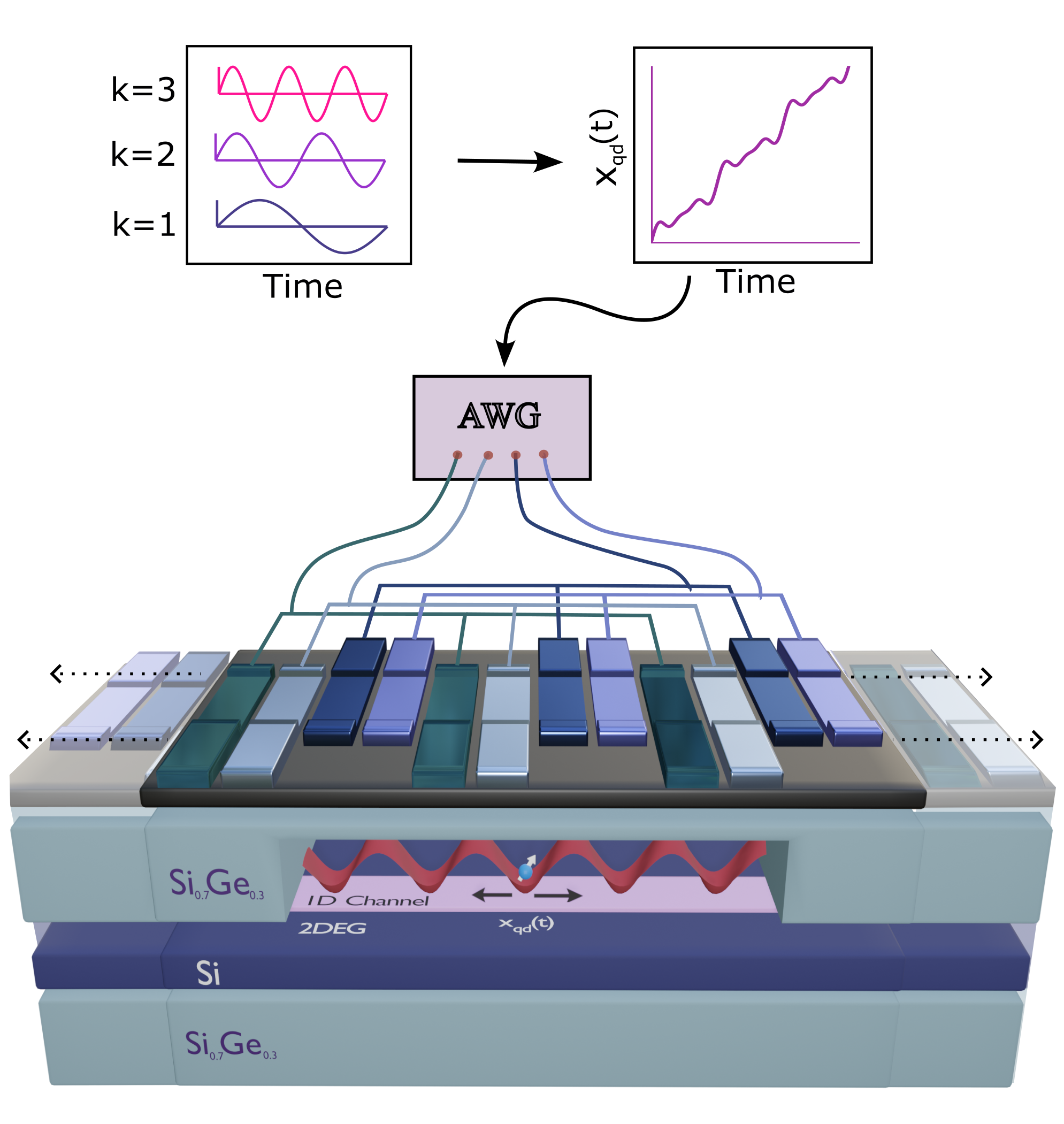}
    \caption{Schematic illustration of an optimized spin transport. Harmonic components are weighted and summed to the constant speed part in order to obtain a modulated speed. We simulate the spin dynamics and we optimized the weighting coefficients as the carrier-hosting quantum dot moves through the valley environment.}
    \label{fig:illustration}
\end{figure}

We consider a single electron trapped in a Si/SiGe quantum dot (QD) at position $x_\text{qd}$, see Fig.~\ref{fig:illustration}. The electron is confined to the orbital ground state as the orbital separation is assumed sufficiently large ($\gtrsim$ 1 meV) to avoid orbital excitations throughout the shuttling operation. In the QD reference frame, the electronic state comprises two components. The first component is a two-level system (TLS) called valley \cite{friesen_valley_2007}, inherited from the strained crystalline structure of Si/SiGe quantum wells. The second component is the electron spin, which defines our qubit and is splitted by a uniform magnetic field, $B_z$. Guided by the device design \cite{langrock_blueprint_2023}, we set $B_z$ = 20~mT. 

Together, valley and spin form a four-level system whose total (position-dependent) hamiltonian $H (x_\text{qd})$ is written as
\begin{eqnarray}
    \label{eq:hamiltonian}
        H (x_\text{qd}) =&& H_S + H_V(x_\text{qd}) + H_{VS}(x_\text{qd}), \nonumber\\
        H_S =&& \frac{E_Z}{2} \, \sigma_z,\nonumber \\
        H_V (x_\text{qd}) =&& \Delta_r (x_\text{qd}) \, \tau_x + \Delta_i (x_\text{qd}) \, \tau_y\nonumber\\ =&& \frac{E_V (x_\text{qd})}{2} \, \tilde{\tau}_z (x_\text{qd}),\nonumber \\
        H_{VS} (x_\text{qd}) =&& - \kappa_z \, \tilde{\tau}_z (x_\text{qd}) \otimes \sigma_z,
\end{eqnarray}
where $E_Z = \bar{g} \mu_B B_z$ is the Zeeman energy, $\bar{g} \approx 2$ is the average Landé $g$-factor, $\mu_B$ is the Bohr magneton and $\sigma_z \equiv \mathbbm{1} \otimes \sigma_z$ is the Pauli $Z$ matrix associated with the spin subsystem.

The position-dependent $\Delta_r (x_\text{qd})$ and $\Delta_i (x_\text{qd})$ are, respectively, the real and imaginary parts of the complex intervalley coupling parameter $\Delta (x_\text{qd}) = \Delta_r (x_\text{qd}) + i \Delta_i (x_\text{qd})$. The absolute value of the intervalley coupling determines the valley splitting $E_V (x_\text{qd}) = 2|\Delta (x_\text{qd})|$.
The valley Hamiltonian term, $H_V(x_\text{qd})$, is defined on the basis $\ket{\pm k_0}$ formed by two Bloch states corresponding to the two minima in the conduction band of strained Si at wavenumbers $\pm k_0 = \pm 0.82 \cdot 2 \pi / a_0$, where $a_0$ = 0.543~nm is the Si unit cell length. The $X$ and $Y$ valley Pauli matrices, $\tau_x \equiv \tau_x \otimes \mathbbm{1}$ and $\tau_y \equiv \tau_y \otimes \mathbbm{1}$, act on the corresponding subspace.
We define the linear combination of valley Pauli matrices $\tilde{\tau}_z (x_\text{qd}) = \cos (\varphi_V (x_\text{qd})) \, \tau_x + \sin (\varphi_V (x_\text{qd})) \, \tau_y$ where $\varphi_V (x_\text{qd}) = \mathrm{arg} \, (\Delta (x_\text{qd}))$ is the valley-mixing phase factor. $\tilde{\tau}_z (x_\text{qd})$ acts as an effective $Z$ Pauli matrix on the local (meaning position-dependent) basis of the ground, $\ket{g (x_\text{qd})}$, and excited, $\ket{e (x_\text{qd})}$, states of the valley, $\ket{e, g (x_\text{qd})} = (\ket{+k_0} \pm e^{i \varphi_V (x_\text{qd})} \ket{-k_0}) / \sqrt{2}$.

The interaction term $H_{VS} (x_\text{qd})$ considered in this work emerges from a measurable difference in $g$-factors between ground and excited valley states \cite{kawakami_electrical_2014, veldhorst_spin-orbit_2015, ferdous_valley_2018, ruskov_electron_2018}. We can relate the measured $g$-factor variation $\delta g / g$ to the coupling term $\kappa_z$ by substituting the valley-dependent $g$-factor $g_v (x_\text{qd}) = \bar{g} + \tilde{\tau}_z (x_\text{qd}) \delta g / 2$ into the Zeeman term $H_Z (x_\text{qd}) = g_v (x_\text{qd}) \mu_B B_z \sigma_z / 2 = H_S + H_{VS} (x_\text{qd})$, obtaining
\begin{equation}\label{eq:kappaz}
    \kappa_z = \frac{1}{4} \frac{\delta g}{g} E_Z.
\end{equation}
Typically, $\delta g / g$ is of the order of $10^{-3}$ leading $\kappa_z$ to be on the order of $10^{-3}$ \textmu{}eV. The minus sign in $H_{VS} (x_\text{qd})$ is just the convention used in this work (sign-change of the $g$-factor valley-dependence).

\subsection{Dynamics}

During shuttling, the position of the QD is changed continuously, $x_\text{qd} \equiv x_\text{qd} (t)$, and the Hamiltonian in Eq.~\eqref{eq:hamiltonian} becomes time-dependent. We need to consider open system dynamics as the valley couples to the phonon environment, leading to relaxation \cite{yang_spin-valley_2013, tahan_relaxation_2014, huang_spin_2014}, while the electron spin interacts with the nuclear spin bath, leading to dephasing \cite{chekhovich_nuclear_2013}. We use a fixed relaxation rate for the valley state, whereas we compute the spin dephasing time from the motional narrowing effect derived in Ref.~\cite{langrock_blueprint_2023}. The approximations involved are addressed in detail in Sec.~\ref{sec:discussion}. Instantaneously, the valley decays from excited to ground state, therefore we choose a position-dependent decay operator equivalent to $\tau_{-} = (\tau_x - i \tau_y) / 2$ on the local ground and excited states, given by
\begin{equation}
    \tilde{\tau}_{-} (x_\text{qd}) = \frac{1}{2} \begin{pmatrix}
        1 & e^{-i \varphi_V (x_\text{qd})} \\
        -e^{i \varphi_V (x_\text{qd})} & -1
    \end{pmatrix}.
\end{equation}
At time $t$, the density matrix of the system, $\rho(t)$, evolves according to the master equation
\begin{multline}
    \label{eq:master}
    \frac{\mathrm{d}\rho}{\mathrm{d}t} (t) = -\frac{i}{\hbar} [H \left ( x_\text{qd} (t) \right ), \rho (t)] \\
        + \frac{1}{T_{1,v}} \mathcal{D} \left [ \tilde{\tau}_{-} \left ( x_\text{qd} (t) \right ) \right] (\rho (t))
        + \frac{1}{2T_{\phi,s}} \mathcal{D} [ \sigma_z ] (\rho (t)),
\end{multline}
where we have the dissipative operator $\mathcal{D}[L](\rho) = L \rho L^\dagger - (\rho L^\dagger L + L^\dagger L \rho) / 2$ and we have used the implicit notation $\tilde{\tau}_{-} (x) \equiv \tilde{\tau}_{-} (x) \otimes \mathbbm{1}$. The relaxation time of the valley is set by $T_{1,v}$, while $T_{\phi,s}$ is the time of pure dephasing for the spin.

\subsection{Valley model}

Accurate simulation of the dynamics requires the assignment of a representative function to the complex parameter $\Delta (x_\text{qd})$ which associates the position of the QD to the strength and phase of the local valley environment covered by the electron wavefunction. There are several models for $\Delta (x_\text{qd})$ which we can adapt to obtain values for the whole length of the device. Currently, the most sophisticated model computes the intervalley coupling from the local atomic arrangement \cite{paquelet_wuetz_atomic_2022, klos_atomistic_2024} and it is consistent with the few experimental results \cite{volmer_mapping_2024}. The empirical coverage of parameter space is not yet complete enough to confirm that this model gives a full picture, as, e.g., strain may also contribute to enhancing the valley splitting \cite{woods_coupling_2024}, but the statistical agreement suggests one obtains a realistic distribution of intervalley couplings.

In effective mass (EM) theory, the intervalley coupling is defined by the integral along the growth direction $\Delta (x_\text{qd}) = \int e^{-2i k_0 z} U(z \, ; x_\text{qd}) |\psi_\perp (z \, ; x_\text{qd})|^2 \, \mathrm{d}z$ \cite{friesen_valley_2007}. Here $U(z \, ; x_\text{qd})$ is the quantum well confinement potential and $\psi_\perp (z \, ; x_\text{qd})$ is the out-of-plane envelope function of the electron. The explicit $x_\text{qd}$ dependence in $U$ and $\psi_\perp$ comes from the atomic arrangement being weighted around the center of the QD by the probability distribution of the in-plane envelope function $\psi_\parallel (x, y \, ; x_\text{qd})$. Indeed, the envelope function that shapes the QD electron wavefunction is separable into out-of-plane and in-plane components, $\Psi (x, y, z \, ; x_\text{qd}) = \psi_\parallel (x, y \, ; x_\text{qd}) \psi_\perp (z \, ; x_\text{qd})$, where $\psi_\parallel (x, y \, ; x_\text{qd}) = \exp \{ -[(x - x_\text{qd})^2 + y^2] / 4 \sigma_\text{qd} \} / (\sigma_\text{qd} \sqrt{2 \pi})$. The characteristic size of the QD, which determines the correlation length of the valley environment, is the standard deviation of the in-plane envelope function squared, estimated as $\sigma_\text{qd} \approx$ 12 nm (see Appendix~\ref{sec:QDradius}). The values of $\Delta (x_\text{qd})$ are sampled every 1.5 nm (justified in Appendix~\ref{sec:QDradius}) and they are interpolated with a cubic spline for the shuttling simulation, in order to avoid jumps in $\mathrm{d}\varphi_V / \mathrm{d}x_\text{qd}$.

The alloy diffusion model is adapted from the theoretical model exposed in Ref.~\cite{paquelet_wuetz_atomic_2022} which is, in turn, based on the effective mass theory of Ref.~\cite{friesen_valley_2007} and the two-band tight binding model of Ref.~\cite{boykin_valley_2004}. This model addresses the influence of Si concentration fluctuations near the well interface to the intervalley coupling. Following Ref.~\cite{paquelet_wuetz_atomic_2022}, the confinement potential $U$ is computed layer by layer by first assigning an average Si concentration profile $0 \le \bar{\xi}(z_l) \le 1$, with $z_l$ the $l$-th layer coordinate. The actual Si concentration is computed by averaging the Si atoms per layer with the in-plane envelope function, $\xi(z_l \, ; x_\text{qd}) = \int \rho_\text{Si}(x, y \, ; z_l) \psi_\parallel (x, y \, ; x_\text{qd}) \, \mathrm{d}x \mathrm{d}y$, where the Si layer density is a sum of Dirac delta functions centered on the Si atom positions $\rho_\text{Si}(x, y \, ; z_l) = \sum_{j \in \text{\{Si\}}} \delta (x - x_j) \delta (y - y_j)$. Each crystal point $(x_j, y_j, z_l)$ is occupied by a Si atom with probability $p = \bar{\xi} (x_l)$ given by the average Si concentration of layer $l$. Here we extend the use of this model by assigning the type of atom (Si or Ge) for a crystal covering the entire 10~\textmu{}m length of the device. Then, for each sampled position $x_\text{qd}$, the crystal is sliced from $x_\text{qd} - 3\sigma_\text{qd}$ to $x_\text{qd} + 3\sigma_\text{qd}$ and this subregion is used to compute $\xi(z_l \, ; x_\text{qd})$. This allows to generate intervalley couplings that share the same atom arrangement and are correlated at adjacent points. Finally, $U(z_l \, ; x_\text{qd})$ is the linear interpolation of the conduction band offset between Si and Si$_{\xi_s}$Ge$_{1-{\xi_s}}$ using the Si concentration $\xi(z_l \, ; x_\text{qd})$ as interpolation parameter (typically the substrate concentration is $\xi_s \sim$ 0.7).

\section{Optimization}

Operating the spin shuttler at constant speed leads to excited valley population and spin decoherence with high probability, as shown in the next section. Here we explore the possibility to alter the spin-valley dynamics to preserve the spin state by changing the trajectory $x_\text{qd} (t)$ of the quantum dot along the transport direction. To quantify the spin-transport infidelity, we measure the entanglement fidelity of the transport evolution with comparison to a unitary evolution precessing the spin at frequency $\nu_G = (\bar{g}\mu_B B_z + 2\kappa_z) / h$, where $h$ is the Planck constant. Following Appendix~\ref{sec:gateFidelity}, we use
\begin{equation}\label{eq:shuttlingfidelity}
    F_\text{ent} (\mathcal{V} \circ \mathcal{E}) = \frac{1}{2} + \mathrm{Re} \big [ \bra{0} (\mathcal{V} \circ \mathcal{E})(\ket{+}\!\!\bra{+}) \ket{1} \big ]
\end{equation}
where $\mathcal{E} (\rho_S(0))$ is the quantum operation taking an initial spin state $\rho_S (0)$ with the groundstate of the valley at starting position $x_\text{qd}(0)$, evolving according to Eq.~\eqref{eq:master} and finally taking the partial trace over the valley subspace. Moreover $\mathcal{V} (\rho_S) = V^\dagger \rho_S V$, where $V = \exp (-2 \pi i \nu_G T)$ and $T$ is the total shuttling time. The average gate fidelity is recovered from the standard formula $\bar{F} = (d F_\text{ent} + 1) / (d + 1)$ \cite{horodecki_general_1999, nielsen_simple_2002}, with $d = 2$ in this case.

We control the shape of the trajectory by optimizing in the frequency domain rather than the time domain \cite{motzoi_optimal_2011, sorensen_optimization_2020}. In particular, we choose a correction to the constant speed trajectory expanded on a sine basis (so that the beginning and the end are fixed points) with $M$ frequency components, from the fundamental frequency $\nu_1 = 1 / T = v / L$ to the bandwidth limit $\nu_M = M \nu_1$,
\begin{equation}\label{eq:trajectory}
    x_\text{qd} (t) = v t + \sum_{k=1}^M u_k \sin (2 \pi \nu_k t),
\end{equation}
where $v$ is the average velocity and $L$ is the total length of the device. The parameter space spanned by the coefficient $\{u_k\}_k$ is explored using the L-BFGS-B algorithm \cite{liu_limited_1989, byrd_limited_1995, zhu_algorithm_1997} implemented in SciPy \cite{zhu_algorithm_1997, virtanen_scipy_2020}.
On the numerical side, the trajectory is discretized into piecewise-constant segments in order to solve the quantum dynamics, $\rho(T) = (\prod_{j=N}^{1} \mathcal{L}_j^{(1)} \mathcal{U}_j) (\rho(0))$, where we use Trotter product formula \cite{trotter_product_1959} and $\mathcal{U}_j (\rho) = U_j \rho U_j^\dagger$, $U_j = \exp (-i H(x_\text{qd}((j-1/2)\delta t))\delta t / \hbar)$ is the unitary part of the evolution step, while $\mathcal{L}_j^{(1)} (\rho) = \rho + \delta t \mathcal{D}[\tilde{\tau}_{-}(x_\text{qd}((j - 1/2)\delta t))] (\rho) / T_{1,v}$ is the dissipative part obtained from the master equation with Euler method (shown only for the valley relaxation); both unitary and dissipative parts employ the midpoint method. The size of the time step $\delta t$ is determined heuristically to give comparable final states as with a 10 times smaller time step over a simulation of 10~\textmu{}m. We determine $\delta t = 0.5$ ps for average speeds $v \gtrsim 50$ m/s, $\delta t = 5$ ps for $v \gtrsim 5$ m/s, $\delta t = 10$ ps for $v \lesssim 2$ m/s and $\delta t = 50$ ps for $v \lesssim 0.2$ m/s. 
The typical number of time steps $N$ that are necessary is in the order of millions, while in some cases the number of harmonics $M$ is in the order of thousands. These large optimizations are obtained from a custom code leveraging the acceleration and automatic differentiation of Python's JAX library \cite{bradbury_jax_2018}. Automatic differentiation is particularly useful in this case as it allows to compute the gradient of of the cost function $I = 1 - F_\text{ent}(\mathcal{V} \circ \mathcal{E})$ with respect to the $\{u_k\}_k$ parameters, required by the BFGS algorithm.

\section{Results}\label{sec:results}

To reflect the typical behavior of shuttlers, e.g. all part of a larger quantum processor, we generate 100 different valley environments following the alloy diffusion model. We present our results for constant speed and optimized shuttling in terms of statistical ensemble averages. Each valley environment is for a straight quantum dot channel of length 10~\textmu{}m. We investigate the behaviour of spin shuttling infidelity and valley excited state population over four main parameters: average shuttling speed $\langle\dot{x}_\text{qd}\rangle = v$, valley relaxation time $T_{1,v}$, valley-spin interaction strength $\kappa_z$ and number of frequency components $M$ of the optimized trajectory (or maximum bandwidth $\nu_M$). The valley environments are generated from a 12~nm quantum well width, $4\tau = 0.8$~nm interface width \cite{paquelet_wuetz_atomic_2022}. The Ge concentration is 30\% in the substrate and 0\% inside the well. The electric field in the growth direction is 0.0125~V/nm. The valley environments each present $238 \pm 6$ minima of valley splitting, for an average of a minimum every 42~nm, and the valley splitting distribution is described by $\mu \pm \sigma = (86 \pm 45)$~\textmu{}eV.

\subsection{Constant speed shuttling}

\begin{figure*}
    \centering
    \includegraphics[width=\textwidth]{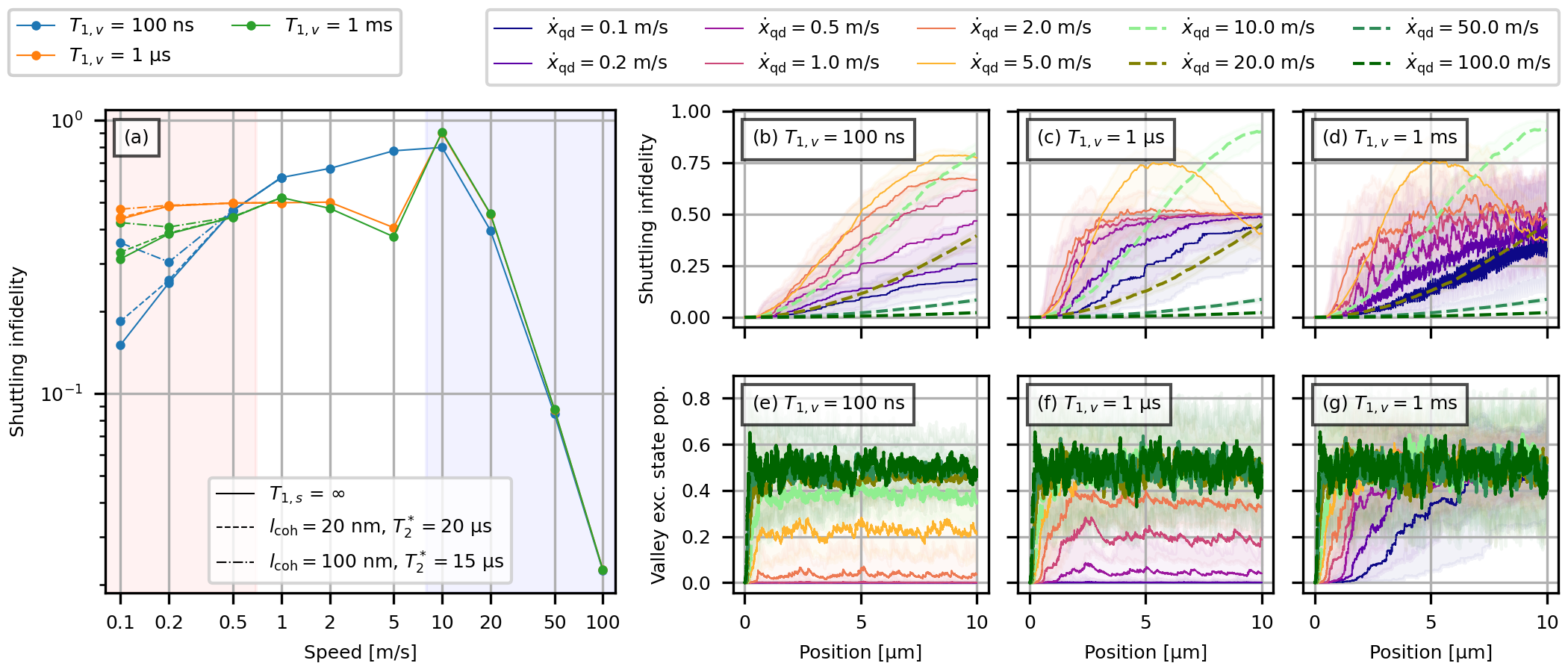}
    \caption{Constant speed shuttling. (a) Spin transport infidelity for various speeds after 10~\textmu{}m of constant speed shuttling (50th percentile of 100 different valley landscape realizations generated by the alloy diffusion model). Different colors indicate different valley lifetimes. Solid lines assume no spin dephasing from slow nuclear spin noise. Dashed lines include slow nuclear noise with coherence length of 20~nm and a $T_2^*$ of 20~\textmu{}s, mitigated by motional narrowing, included as an effective speed-dependent spin dephasing. Dot-dashed lines consider nuclear noise with coherence length of 100~nm and a $T_2^*$ of 15~\textmu{}s. The red-shaded background indicates the region where nuclear noise has a significant contribution to the shuttling infidelity. The blue-shaded background show the effect of higher speed where the valley has progressively less influence. 
    (b--g) Spin transport infidelity (first row) and population of the excited valley state (second row) as a function of the position along the shuttler (nuclear noise with coherence length 20 nm, coherence time 20 \textmu s). The different columns have valley lifetime of $T_{1,v} = 100$ ns (left), $T_{1,v} = 1$ \textmu s (center), $T_{1,v} = 1$ ms (right). Solid lines indicate the 50th percentile of 100 traces obtained from 100 different valley landscape realizations generated by the alloy diffusion model. Shaded area indicate the 25th to 75th percentile range for the corresponding color.}
    \label{fig:multi_plot_const_speed_infid_population}
\end{figure*}

During constant speed shuttling, the quantum dot trajectory used in Eq.~\eqref{eq:master} is $x_\text{qd} (t) = v t$ for $0 \le t \le T = L / v$. Different regimes emerge depending on the shuttling velocity, as highlighted in Fig.~\ref{fig:multi_plot_const_speed_infid_population}(a). Up to 10~m/s the main contribution to the shuttling infidelity comes from the high number of valley splitting minima associated also with a large valley phase gradient.
This induces diabatic transitions in the excited valley state with high probability, leading to different spin precession
rates from the $g$-factor valley-dependence and, therefore, valley-spin entanglement. Furthermore, as the valley relaxes, decoherence is induced on both spin and valley states increasing the shuttling infidelity. 

At low speed, the spin dynamics is also influenced by the slow nuclear spin noise as indicated by the dashed and dash-dotted lines in the red-shaded area. Above 1~m/s the effect of motional narrowing inhibits the dephasing coming from nuclear noise. Calculation of motional narrowing follows the derivation of Ref.~\cite{langrock_blueprint_2023} and we translate the dephasing mechanism in an effective $T_{\phi,s} = v (T_2^*)^2 / (2l_c)$ that we use in Eq.~\eqref{eq:master}, where $T_2^*$ is the spin-dephasing time observed for stationary spins and $l_c$ is the coherence length of the noise sources (see Appendix~\ref{sec:motionalnarrowing}). 

Above 10~m/s the phase rotations from valley-spin coupling have progressively less time to deviate from the rotating frame of the valley ground state. Above this speed the shuttling infidelity decreases monotonically as we reach the (spin) diabatic shuttling regime and the valley has mitigated overall effect, though still large by error correction standards.

In Fig.~\ref{fig:multi_plot_const_speed_infid_population}(b--g) we show the evolution of shuttling infidelity --- first row (b--d) --- and valley excited state population --- second row (e--g) --- as a function of the position of the quantum dot during shuttling. Each column presents a different valley lifetime. Remarkably, panels (e) and (f) indicate that the valley excited population has an average saturation value that depends on shuttling speed and valley relaxation time. This follows from the generally uniform distribution of valley splitting minima; crossing this minima induces excitations with increasing probability at higher speeds similarly to Landau-Zener transitions, but in between minima the valley has time to relax and the process reaches an equilibrium. This phenomenon could be leveraged in experiments to determine the valley relaxation rate or the average distance between valley minima by measuring the average excited valley population at different speeds. To complete this picture, from panel (g) we notice that, as the valley lifetime increases, the valley population has time to build up reaching half filling at all speeds considered. This is the maximum average occupation given by the multiple random scatterings between ground and excited state. This maximum value is nevertheless reached with different rates (slopes) for different speeds.

\begin{figure}
    \centering
    \includegraphics[width=\columnwidth]{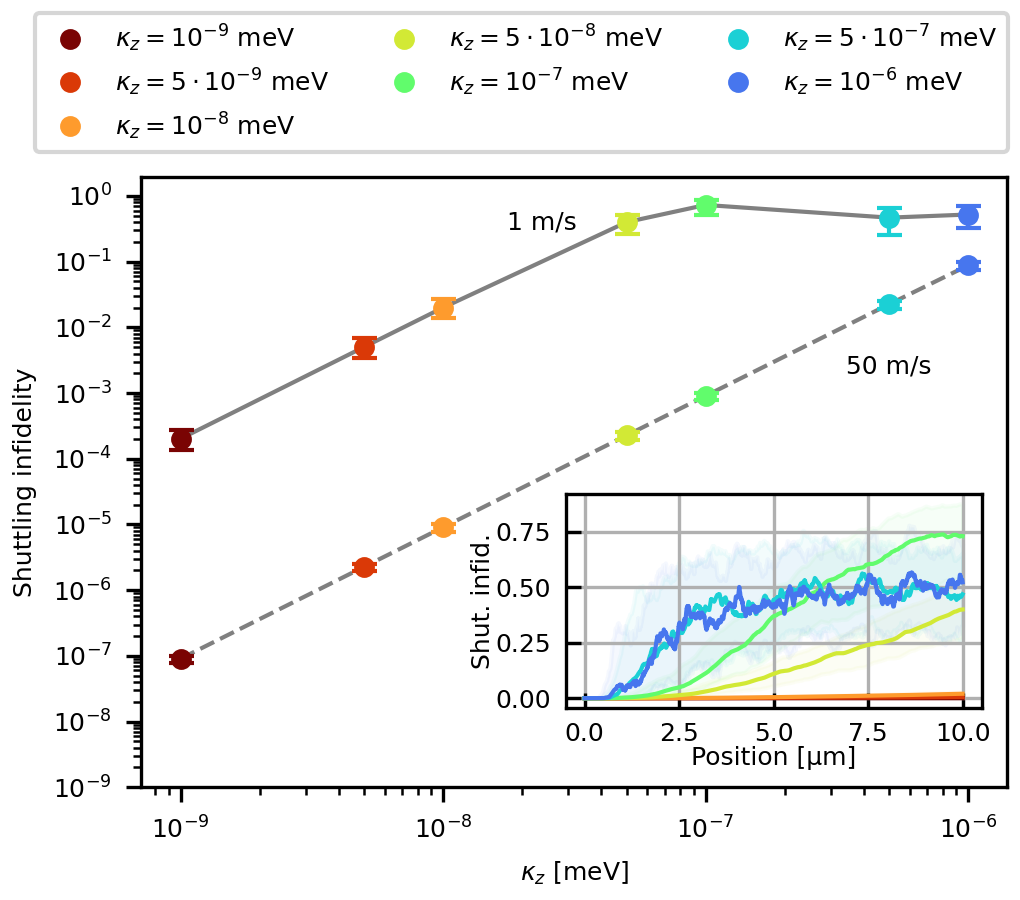}
    \caption{Final shuttling infidelity for constant speed shuttling, with different valley-spin coupling $\kappa_z$. The inset shows the shuttling infidelity as a function of spatial position for 1~m/s speed.}
    \label{fig:const_speed_different_gammas}
\end{figure}

In all the simulations we set the valley-spin coupling $\kappa_z = 10^{-6}$~meV as this is approximately the value obtained from Eq~\eqref{eq:kappaz} for the operational external magnetic field $B_z$ = 20~mT and a $g$-factor variation $\delta g / g$ = 0.1\%. The latter value has been estimated empirically only a few times, spanning an order of magnitude \cite{kawakami_electrical_2014, veldhorst_spin-orbit_2015, ferdous_valley_2018}. Moreover, fluctuations of the $g$-factor variation along the shuttling device are possible and may suppress the shuttling infidelity to fault tolerant values. For this reason we investigate in Fig.~\ref{fig:const_speed_different_gammas} the behaviour of the shuttling infidelity at 1~m/s as well as 50~m/s for different values of $\kappa_z$. For 1~m/s we see that the shuttling infidelity remains above 10\% until $\kappa_z = 5\cdot 10^{-8}$~meV after which it decreases with a power low scaling, reaching fault-tolerant values around $\kappa_z = 10^{-9}$~meV. Instead, for 50~m/s the power low regime is present over all coupling strengths considered and the shuttling infidelity reaches fault-tolerant values below $\kappa_z=10^{-7}$~meV.

\subsection{Optimized shuttling}\label{sec:optimization}

In the optimization process the shuttling fidelity, Eq.~\eqref{eq:entangFidelity}, is maximized exploring the $M$-dimensional space of the $\{u_k\}_k$ coefficients parametrizing the trajectory of Eq.~\eqref{eq:trajectory}.

\begin{figure}
    \centering
    \includegraphics[width=\columnwidth]{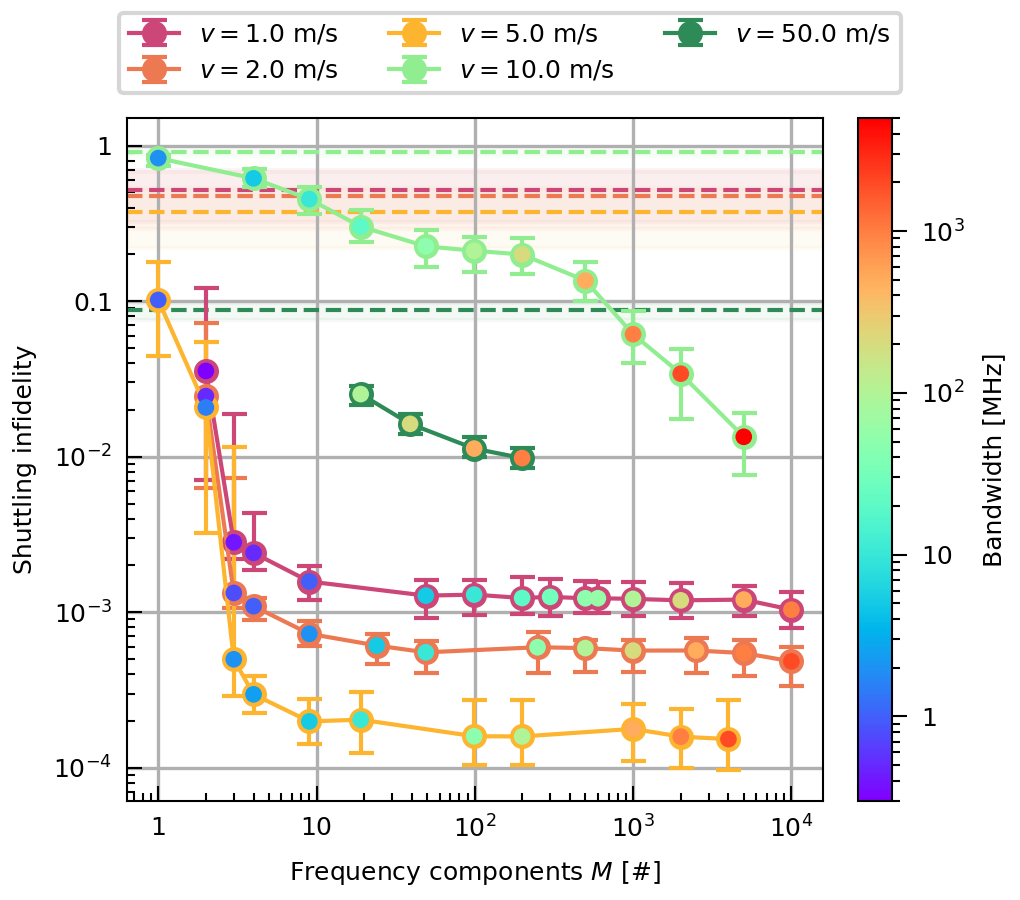}
    \caption{Optimized shuttling infidelity vs numbers of frequency components, $M$, in the correction to constant speed shuttling (Eq.~\eqref{eq:trajectory}); comparison for different speeds. Valley lifetime is 1~ms and valley-spin coupling is $\kappa_z = 10^{-6}$~meV. Each circle on the solid lines represent the 50th percentile of the 100 valley environments considered, while error bars indicate the 25th to 75th interval. The center of each circle is colored according to the colorbar on the right to indicate the maximum bandwidth $\nu_M$, computed as $\nu_M = M v / L$. The dashed lines indicate the 50th percentile of the constant speed shuttling for the corresponding speed color and the shaded background is the 25th to 75th interval.}
    \label{fig:optimization_different_bandwidths}
\end{figure}

The main result is shown in Fig.~\ref{fig:optimization_different_bandwidths} where we plot the minimized shuttling infidelity in function of the frequency components included in the trajectory for five different speeds ($v$ = 1, 2, 5, 10, 50~m/s). This is done for valley lifetime of 1~ms and coupling strength $\kappa_z = 10^{-6}$~meV. The common behaviour for speeds up to 5~m/s is to approach a plateau above 9 frequency components. The infidelity at the plateaus decrease in value as the speed increases from 1 to 5~m/s, reaching an infidelity of $1-F_\text{ent} = 2 \cdot 10^{-4}$ at 5~m/s which is 3 orders of magnitude lower than the constant speed shuttling result indicated by the dashed line of the corresponding color. This is obtained, e.g., for as low as 9 frequency components, but 3 and 4 components also perform remarkably well. At 2~m/s we observe an infidelity of $10^{-3}$ with 4 components. For 10 and 50~m/s the resulting infidelities are above the fault-tolerant threshold, with values monotonically decreasing with more harmonics. Infidelities for 50~m/s are visibly better than for 10~m/s reflecting the situation at constant speed where the shorter time of interaction between spin and valley leads already to lower infidelities for 50~m/s.

\begin{figure}
    \centering
    \includegraphics[width=\columnwidth]{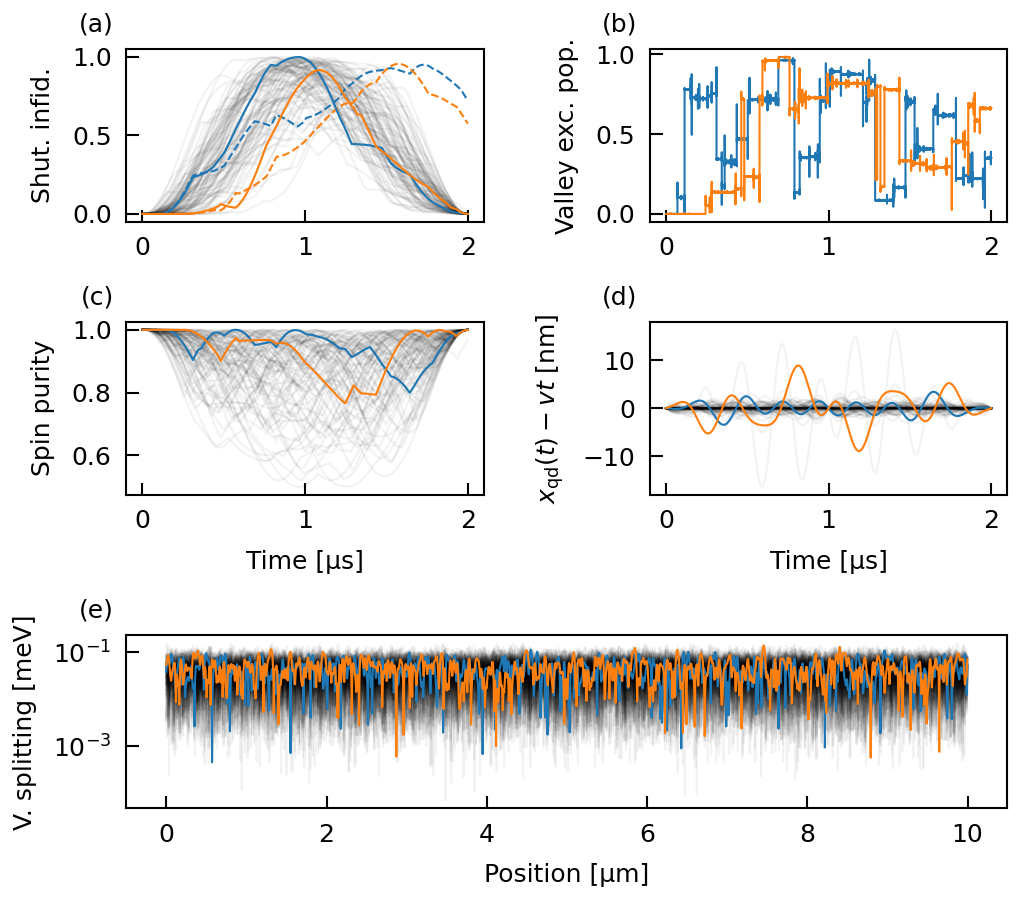}
    \caption{Key quantities during shuttling for the optimized trajectories at 5~m/s and 9 frequency components (bandwidth of 5 MHz). The valley lifetime is 1~ms. The different traces (transparent gray lines) correspond to the 100 valley environments considered, where two particular traces are highlighted (blue and orange solid lines). (a) Shuttling infidelity vs time; the dashed line are the constant speed infidelities for the same valley environments. (b) Valley population during (optimized) shuttling for the two highlighted environments. (c) Spin purity vs time. (d) Optimized correction to the constant speed trajectory. (e) Valley splitting along the shuttler device.}
    \label{fig:multiplot_optimized_5_m_s}
\end{figure}

We complement this picture showcasing in Fig.~\ref{fig:multiplot_optimized_5_m_s} the optimized outcomes for speed of 5~m/s and 9 frequency components. In Fig.~\ref{fig:multiplot_optimized_5_m_s}(a), we plot the 100 optimized shuttling infidelities, coming from the different valley environments, as a function of time during shuttling. We also highlight two of these optimized traces (blue and orange solid lines) and we compared them to their constant speed shuttling counterparts coming from the same valley environments (dashed lines). We observe a common behavior in the optimized traces with high infidelity in the middle and back to low infidelity at the end. This is reflected in the spin purity plotted in Fig.~\ref{fig:multiplot_optimized_5_m_s}(c) where the central part of shuttling is characterized by low spin purity, indicating entanglement with valley, before returning to a pure state at the end. This indicates that in this regime the optimizer does not try to avoid shuttling infidelity, because it does not have high enough frequencies to keep the valley in the ground state. Rather, it takes advantage of the valley-spin coupling to let the spin perform a complete rotation. For a valley-spin coupling of $\kappa_z = 10^{-6}$~meV the rotation frequency at 50\% valley population would be $2\kappa_z/h \approx 0.242$~MHz corresponding to a period of rotation of 2.07~\textmu{}s. This is close to the total shuttling time at 5~m/s and it explain why the optimization for 10~m/s performs suddenly worse, i.e., the spin has time only for a half rotation. Therefore, we remark that the shuttling speed giving the best result after optimization will depend on the interaction strength $\kappa_z$ between spin and valley and the shuttling distance.

\begin{figure}
    \centering
    \includegraphics[width=\columnwidth]{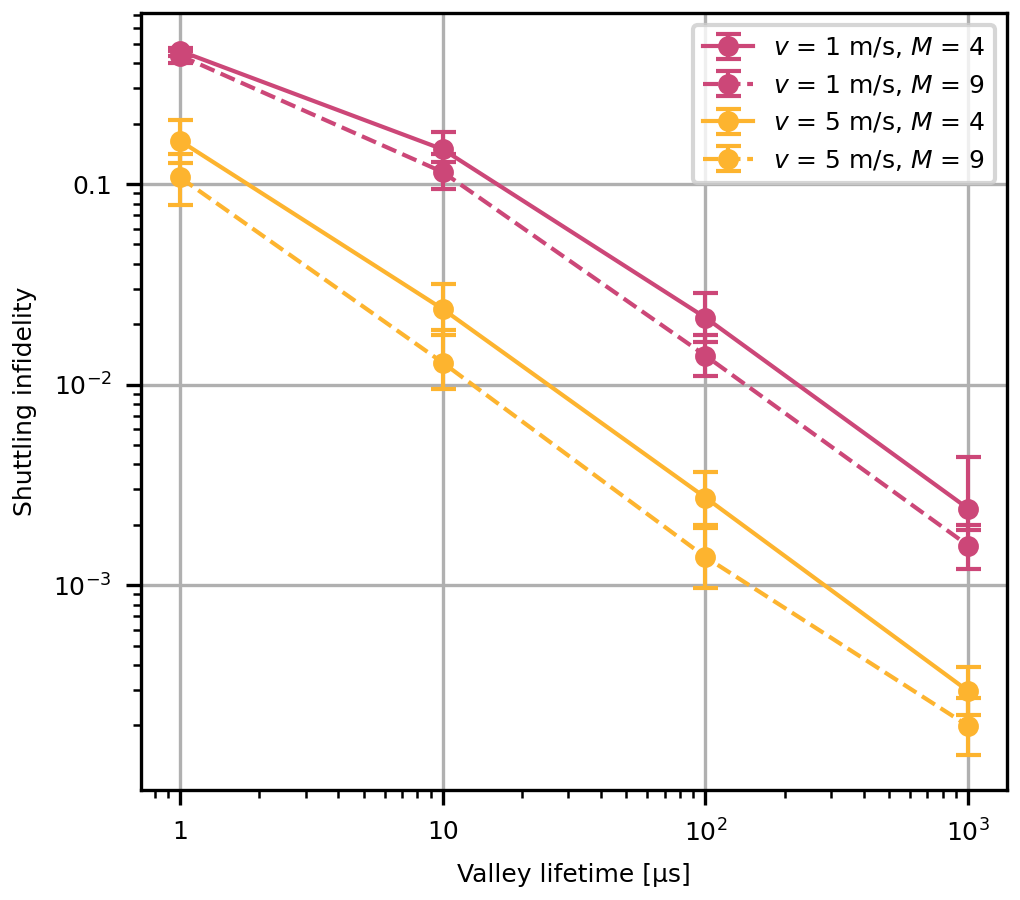}
    \caption{Optimized shuttling, comparison of shuttling infidelity for different valley lifetimes at 1~m/s (red/purple) and at 5~m/s (yellow/orange). Circles indicate the 50th percentile while error bars indicate the 25th to 75th percentile of optimized values for the 100 valley environments considered.}
    \label{fig:optimization_different_valley_lifetimes}
\end{figure}

Finally, we consider the effect of valley lifetime on the spin transport optimization. In Fig.~\ref{fig:optimization_different_valley_lifetimes} we plot the statistics of the final shuttling infidelity for valley lifetimes ranging from 1~\textmu{}s to 1~ms and for an average shuttling speed of 1~m/s and 5~m/s. We compare two different number of components, 4 and 9. We observe a monotonic decrease of the infidelity from 1~\textmu{}s lifetime to 1~ms, with an approximate power law behaviour, but only the longest lifetimes reach fault-tolerant values. Overall, the right range of values for the valley lifetime has not been yet experimentally determined and it should be a function of the local valley splitting, requiring more advanced models for accurate simulations.

\subsection{Discussion}\label{sec:discussion}

In the model used to describe the shuttling process we have introduced a number of approximations. These approximations simplify the analysis of the influence of the valley decay on the shuttling infidelities while keeping predictive power.

Regarding valley relaxation, we have used a fixed relaxation rate, although there is in principle a $E_V^5$ dependence \cite{yang_spin-valley_2013, tahan_relaxation_2014} with the valley splitting. Many details of this dependence are not empirically known, e.g. the normalization constant or the influence of local strain fluctuations, therefore we have used an effective description of the phenomenon.
We let the valley lifetime, $T_{1,v}$, span several orders of magnitude, compatibly with available values from the literature. Direct measurement indicates a 12~ms valley lifetime for a valley splitting of 40~\textmu{}eV \cite{penthorn_direct_2020}. Indirect measurements are provided by the spin relaxation rates at the spin-valley hotspots \cite{yang_spin-valley_2013, borjans_single-spin_2019, hollmann_large_2020}. The data points at the hotspot are between 100~\textmu{}s and 1~ms, as reported for example by Ref.~\cite{hollmann_large_2020} for a valley splitting above 200~\textmu{}eV. Our valley environments generated with the alloy diffusion model have a valley splitting below 200~\textmu{}eV in 98.5\% of the positions. These considerations point to conclude that on average the valley lifetime should be around 1~ms or above, which is also the range where our optimization technique works better, as outline by Fig.~\ref{fig:optimization_different_valley_lifetimes}.

Dephasing between local ground and excited valley states is naturally present. Coupling between spin and valley at the hotspot can produce coherent oscillations for about 2~\textmu{}s before fading \cite{cai_coherent_2023}. In our case, away from the hotspot, the interaction between the spin and valley eigenbases is of $ZZ$-type and the valley phase does not affect the spin-subspace. Therefore the phase accumulated on the spin by the interaction would not be influenced by valley dephasing, but only from the relative valley populations in the ground and excited states. Nevertheless, valley dephasing would influence the valley populations as the local eigenbasis changes during the QD transport. The effect of valley dephasing on spin shuttling needs further investigating.

The spin-valley hybridization at the hotspot, mentioned above, induces faster relaxation for the spin. Here we have not included this mechanism and in Appendix~\ref{sec:hotspotRelaxation} we quantify its negligible contribution.

\section{Conclusions}\label{sec:Conclusions}

Through our simulations, we identify two methods to reduce the spin shuttling infidelity. In the constant speed shuttling regime our results indicate that speeds higher than 50~m/s lead to low infidelity values by decreasing the interaction time between spin and valley degrees of freedom but still well above fault tolerance requirements. On the other hand, better results are possible  when using an optimization scheme. For optimal results, the shuttling time should be close to the period of spin precession determined by the valley-spin interaction.
 
We suggest a simple parametrization of the quantum dot trajectory by expanding the correction to constant speed shuttling into harmonic components. Infidelity values lower than $10^{-3}$ are reached for a number of frequency components between 4 and 20. At the end of the optimized shuttling the spin is disentangled from the valley and it can be used in subsequent single- and two-qubit operations. The low number of frequency components also allows gradient-free optimization methods such as Nelder-Mead. This number is a lot lower than would be required for full manipulation of the valley state, which would scale with the length of the channel. Instead the procedure we describe here avoids the need for precise spatial information about valley crossings. Our only assumptions for this scheme is that the shuttling trajectory can be realized with sufficient accuracy and that only one spin carrier is shuttled at a time. Other device parameters match current experimental values, therefore we deem that our approach is ready to be experimentally tested.

\begin{acknowledgments}

We thank Veit Langrock, Max Oberländer, Michael Schilling and Francesco Preti for useful discussions. Funded by the Deutsche Forschungsgemeinschaft (DFG, German Research Foundation) under Germany's Excellence Strategy – Cluster of Excellence Matter and Light for Quantum Computing (ML4Q) EXC 2004/1 – 390534769.

\end{acknowledgments}

\appendix

\section{QD radius and intervalley coupling sampling}\label{sec:QDradius}

In a conveyor-mode shuttling device like the proposed QuBus architecture \cite{langrock_blueprint_2023} the (qubit) charge carrier sits at the bottom of a sinusoidal potential in the $x$-direction and of a convex potential in the $y$-direction. We approximate the bottom of this potential with a 2D isotropic harmonic trap with orbital separation $E_0 = \hbar\omega_h$. The in-plane envelope function $\psi_\parallel (x, y \, ; x_\text{qd})$ corresponds to the 2D Gaussian ground state and we identify the characteristic size of the dot with the standard deviation of the probability distribution (amplitude squared) obtained as
\begin{equation}
    \sigma_\text{qd} = \sqrt{\frac{\hbar}{2 m^* \omega_h}} = \frac{\hbar}{\sqrt{2 m^* E_0}} \approx \frac{14.16 \, [\text{nm} \sqrt{\text{meV}}]}{\sqrt{E_0 \, [\text{meV}]}},
\end{equation}
where $m^* = 0.19 m_e$ is the effective mass of an electron moving in-plane in Si/SiGe. We set $\sigma_\text{qd}$ = 12 nm which corresponds to $E_0$ = 1.4 meV, satisfying the device operation requirements. The orbital separation can also be estimated from the device's gate pitch $p_g$ as the sinusoidal potential has wavelength $\lambda = 4p_g$. Approximating the potential $V_s (x) = -V_0 \cos (k_s x)$ with a harmonic potential $V_h (x) = k_h x^2 / 2$ around $x = 0$ one obtains $k_h = V_0 k_s^2$ and therefore $m^* \omega_h^2 = V_0 (2\pi / \lambda)^2$.

The characteristic size $\sigma_\text{qd}$ determines the correlation length of the valley environment. This has to be taken into account when choosing the sampling distance $d$ for the intervalley coupling $\Delta(x_\text{qd})$. Consider two identical 2D Gaussian functions with standard deviation $\sigma_\text{qd}$ and normalized to 1. At an in-plane distance $d$ from each other, the overlap volume is
\begin{equation}
    V_\text{overlap} = 1 - \mathrm{erf} \left ( \frac{d}{2\sqrt{2}\,\sigma_\text{qd}} \right ),
\end{equation}
where $\mathrm{erf}$ is the error function. If we set the sampling distance $d$ = 1.5~nm when $\sigma_\text{qd}$ = 12~nm, we ensure that two adjacent intervalley couplings share 95\% of the same atomic environment. Arbitrary $\Delta(x_\text{qd})$ can then be obtained by smooth interpolation.

\section{Gate fidelity for Z-commuting evolution}\label{sec:gateFidelity}

We use the entanglement fidelity \cite{} defined as
\begin{equation}\label{eq:entangFidelity}
    F_\text{ent} (\mathcal{E}) = \bra{\phi} (\mathbbm{1} \otimes \mathcal{E}) (\phi) \ket{\phi}
\end{equation}
where the quantum channel $\mathcal{E}$ acts on system $S$ and $\ket{\phi}$ is a maximally entangled state of the joint system $A \,\otimes\, S$ with $A$ as an ancilla system of the same size of $S$. For a one-qubit channel we use $\ket{\phi} = (\ket{00} + \ket{11}) / \sqrt{2}$ and we expand Eq.~\eqref{eq:entangFidelity} to obtain
\begin{multline}
    F_\text{ent}(\mathcal{E}) = \frac{1}{4} 
    \big ( \bra{0} \mathcal{E} (\ket{0}\!\!\bra{0}) \ket{0} + 
    \bra{0} \mathcal{E} (\ket{0}\!\!\bra{1}) \ket{1} + \\
    \bra{1} \mathcal{E} (\ket{1}\!\!\bra{0}) \ket{0} + 
    \bra{1} \mathcal{E} (\ket{1}\!\!\bra{1}) \ket{1} \big ).
\end{multline}
Since $\ket{0}\!\!\bra{1}$ and $\ket{1}\!\!\bra{0}$ are not physical states, one usually substitutes them with linear combinations of other states like
\begin{equation}
    \begin{aligned}
        \ket{0}\!\!\bra{1} & = \ket{+}\!\!\bra{+} + i \ket{+i}\!\!\bra{+i} - (1+i) (\ket{0}\!\!\bra{0} + \ket{1}\!\!\bra{1}) / 2 \\
        \ket{1}\!\!\bra{0} & = \ket{+}\!\!\bra{+} - i \ket{+i}\!\!\bra{+i} - (1-i) (\ket{0}\!\!\bra{0} + \ket{1}\!\!\bra{1}) / 2
    \end{aligned}
\end{equation}
where $\ket{+} = (\ket{0} + \ket{1}) / \sqrt{2}$ and $\ket{+i} = (\ket{0} +i \ket{1}) / \sqrt{2}$. Finally, using the linearity of $\mathcal{E}$, we compute the entanglement fidelity as
\begin{multline}\label{eq:qubitEntangFidelity}
    F_\text{ent} (\mathcal{E}) = \frac{1}{4} \big [ \bra{0} \mathcal{E}_0 \ket{0} + \bra{1} \mathcal{E}_1 \ket{1} \\
    + \bra{0} \mathcal{E}_+ \ket{1} + i \bra{0} \mathcal{E}_{+i} \ket{1} - (1 + i) \bra{0} \mathcal{E}_\mathbbm{1} \ket{1} \\
    + \bra{1} \mathcal{E}_+ \ket{0} - i \bra{1} \mathcal{E}_{+i} \ket{0} - (1 - i) \bra{1} \mathcal{E}_\mathbbm{1} \ket{0} \big ]
\end{multline}
from the evolution of 4 physical states, where $\mathcal{E}_0 = \mathcal{E}(\ket{0}\!\!\bra{0})$, $\mathcal{E}_1 = \mathcal{E}(\ket{1}\!\!\bra{1})$, $\mathcal{E}_+ = \mathcal{E}(\ket{+}\!\!\bra{+})$, $\mathcal{E}_{+i} = \mathcal{E}(\ket{+i}\!\!\bra{+i})$ and $\mathcal{E}_\mathbbm{1} = \mathcal{E}(\mathbbm{1}/2) = (\mathcal{E}_0 + \mathcal{E}_1) / 2$.

Consider a one-qubit evolution $\mathcal{E}$ generated by a master equation in Lindblad form and an operator $U$ which commutes with the Hamiltonian and the jump operators $L_k$ as well as the operators $L_k^\dagger L_k$, $\forall k$. Then we have the property $\mathcal{E} (U \rho U^\dagger) = U \mathcal{E} (\rho) U^\dagger$ (alternatively $U$ should commute with all the Kraus operator of a give Kraus decomposition of $\mathcal{E}$). In our case, Eq.~\eqref{eq:master} has such symmetry for $U = \sigma_z$ as well as all unitary operators generated by $\sigma_z$. Then we can further simplify the expression above considering $\mathcal{E}_0 = \ket{0}\!\!\bra{0}$, $\mathcal{E}_1 = \ket{1}\!\!\bra{1}$ and $\mathcal{E}(\ket{+i}\!\!\bra{+i}) = \mathcal{E}(V\ket{+}\!\!\bra{+}V^\dagger) = V \mathcal{E}(\ket{+}\!\!\bra{+}) V^\dagger$, with $V = \ket{+i}\!\!\bra{+} + \ket{-i}\!\!\bra{-}$ (generated by $\sigma_z$), therefore giving
\begin{equation}\label{eq:zsymmetryfidelity}
    \begin{aligned}
        F_\text{ent} (\mathcal{E}) 
        & = \frac{1}{4} \big [ 2 + 2 \bra{0} \mathcal{E}_+ \ket{1} + 2 \bra{1} \mathcal{E}_+ \ket{0} \big ] \\
        & = \frac{1}{2} + \mathrm{Re} \big ( \bra{0} \mathcal{E}_+ \ket{1} \big )
    \end{aligned}
\end{equation}

\section{Motional narrowing}\label{sec:motionalnarrowing}

Here we approximate the effect of quasistatic noise on the electron spin with a pure dephasing evolution, taking into account the effect of motional narrowing described in Ref.~\cite{langrock_blueprint_2023}. The pure dephasing part of our Lindblad master equation (Eq.~\eqref{eq:master}) reads
\begin{equation}
    \dot{\rho} = \frac{1}{2T_{\phi,2}}(\sigma_z \rho \sigma_z - \rho),
\end{equation}
which gives the solution
\begin{equation}\label{eq:dephasingsolution}
        \bra{0}\rho (t)\ket{1} = \bra{0}\rho (0)\ket{1} e^{-t/T_{\phi,s}}
\end{equation}
and equivalently for the other off-diagonal element. This process describes a purely dephasing channel $\mathcal{E}_\phi$ for the duration of the shuttling time $T = L/v$. Plugging Eq.~\eqref{eq:dephasingsolution} in Eq.~\eqref{eq:zsymmetryfidelity} we obtain
\begin{equation}
    \begin{aligned}
    F_\text{ent} (\mathcal{E}_\phi) & = \frac{1}{2} + \mathrm{Re} \big ( \braket{0 | +} \! \braket{+ | 1} \big ) e^{-T/T_{\phi,s}} \\
    & = \frac{1}{2} \big ( 1 + e^{-T/T_{\phi,s}} \big ).
    \end{aligned}
\end{equation}
As explained in Ref.~\cite{losert_strategies_2024}, a gaussianly distributed dephasing noise with rms $\delta \Phi$ gives an expected infidelity $\langle I \rangle = 1 - \langle F_\text{ent} \rangle \approx \delta \Phi^2 / 4$. In Ref.~\cite{langrock_blueprint_2023}, the rms of quasistatic noise including the motional narrowing effect is estimated to be $\delta \Phi^2 = 4 l_c L / (v T_2^*)^2$, where $l_c$ is the coherence length of the noise sources and $T_2^*$ is the spin-dephasing time observed for stationary spins. Comparing this estimate with the dephasing channel infidelity in the limit of large $T_{\phi,s}$, we have
\begin{equation}
    I = 1 - F_\text{ent} (\mathcal{E}_\phi) \approx \frac{T}{2T_{\phi,s}} \equiv \frac{l_c L}{(v T_2^*)^2} = \frac{l_c T}{v (T_2^*)^2},
\end{equation}
finally arriving to
\begin{equation}
    T_{\phi,s} \equiv \frac{v}{2l_c} (T_2^*)^2.
\end{equation}

\section{Spin relaxation at the hotspot}\label{sec:hotspotRelaxation}

When the valley splitting is close to the spin splitting, the hybridization between spin and valley mediated by spin-orbit coupling leads to faster relaxation rates for the spin \cite{yang_spin-valley_2013, huang_spin_2014, borjans_single-spin_2019, hollmann_large_2020}. In our model we have neglected this effect and here we show that its error contribution is below the harming threshold. At low spin splitting and away from possible hotspots, spin relaxation times are limited to 1~s by Johnson noise \cite{hollmann_large_2020}. For the timescales involved in this work we consider the spin lifetime away from the hotspot as infinite. The rate of spin decay can be expressed as a function of the distance between spin splitting and valley splitting as \cite{yang_spin-valley_2013}
\begin{equation}
    \Gamma_s (\delta) = \left ( 1 - \frac{|\delta|}{\sqrt{\delta^2 + \Delta_\text{so}^2}}\right ) \frac{\Gamma_v}{2},
\end{equation}
where $\delta = E_V - E_S$ with $E_V$ and $E_S$, respectively, the valley and spin splitting, while $\Gamma_v$ is the valley decay rate. The strength of the spin-valley mixing $\Delta_\text{so}$ plays a role determining the width of the spin relaxation rate spike at the hotspot. When spin and valley splitting coincide, the spin decay rate is half the valley decay rate, $\Gamma_s(0) = \Gamma_v / 2$, while when $\delta = \Delta_\text{so}$, the spin decay rate drops by about 20\%, $\Gamma_s (\Delta_\text{so}) = (\sqrt{2}-1) \Gamma_v / 4 \approx 0.207 \, \Gamma / 2$. From this and looking at, e.g., Ref.~\cite{borjans_single-spin_2019}, we estimate $\Delta_\text{so} \approx$ 6~\textmu{}eV at $E_V =$ 106~\textmu{}eV. Of course $\Delta_\text{so}$ (as well as $\Gamma_v$) decreases at lower values of $E_V$ \cite{huang_spin_2014}.

The relaxation mechanism for the spin state $\rho$ follows the master equation
\begin{equation}
    \dot{\rho} = \Gamma_s (\sigma_- \rho \sigma_+ - \sigma_+ \sigma_- \rho - \rho \sigma_+ \sigma_-),
\end{equation}
where we have used $\sigma_\pm = (\sigma_x \pm i \sigma_y) / 2$. The evolved state $\rho (\Delta t)$ after a time step $\Delta t$ spent at a decay rate $\Gamma_s \equiv \Gamma_s^{(1)}$ is
\begin{equation}\label{eq:relaxationtimestep}
    \rho (\Delta t) = \begin{pmatrix}
        \rho_{00} (0) e^{-\Gamma_s^{(1)}\Delta t} & \rho_{01} (0) e^{-\Gamma_s^{(1)}\Delta t / 2} \\
        \rho_{10} (0) e^{-\Gamma_s^{(1)}\Delta t / 2} & 1-\rho_{00} (0) e^{-\Gamma_s^{(1)}\Delta t}
    \end{pmatrix},
\end{equation}
where $\rho_{jk}(0)$ are the initial conditions. After $n$ time steps spent at decay rates, respectively, $\Gamma_s^{(1)}, \Gamma_s^{(2)}, \ldots, \Gamma_s^{(n)}$, we can substitute $\Gamma_s^{(1)}$ in Eq.~\eqref{eq:relaxationtimestep} with the sum $\Gamma_s^{(1)} + \cdots + \Gamma_s^{(n)}$. Each decay rate $\Gamma_s^{(j)}$ is computed from a different valley splitting $E_V^{(j)}$ at position $x_\text{qd}(j\Delta t)$.

The fidelity of the above evolution can be computed by applying Eq.~\eqref{eq:qubitEntangFidelity}, which gives
\begin{multline}
    F_\text{ent}(\rho(0) \mapsto \rho(n\Delta t)) = \\ \frac{1}{4} \left \{ 1 + \exp \big [-(\Gamma_s^{(1)} + \cdots + \Gamma_s^{(n)})\Delta t / 2 \big ] \right \}^2.
\end{multline}
We use $E_S$ = 2.3~\textmu{}eV (corresponding to $B_z$ = 20~mT) and we keep $\Gamma_v$ = 1/(100~\textmu{}s), $\Delta_\text{so}$ = 6~\textmu{}eV, which are conservatively high values for $E_V \approx E_S$. For our 100 valley environments, we obtain on average $1-F_\text{ent}$ = $3.3 \cdot 10^{-4}$ for a 10~\textmu{}m shuttling at average speed of 1~m/s. Increasing the speed to 5~m/s we have $1-F_\text{ent}$ = $6.7 \cdot 10^{-5}$. Moreover, decreasing by an order of magnitude both $\Gamma_v$ and $\Delta_\text{so}$, which is expected, we obtain $1-F_\text{ent}$ = $1.6 \cdot 10^{-6}$ at 1~m/s. Therefore, we predict that spin relaxation originating from the spin-valley hybridization does not contribute significantly to the shuttling infidelity for $B_z$ = 20~mT.

\bibliography{bibliography.bib}

\widetext

\newpage

\begin{center}
\textbf{\large Supplementary Information}
\end{center}

\begin{figure*}
    \centering
    \includegraphics[width=\textwidth]{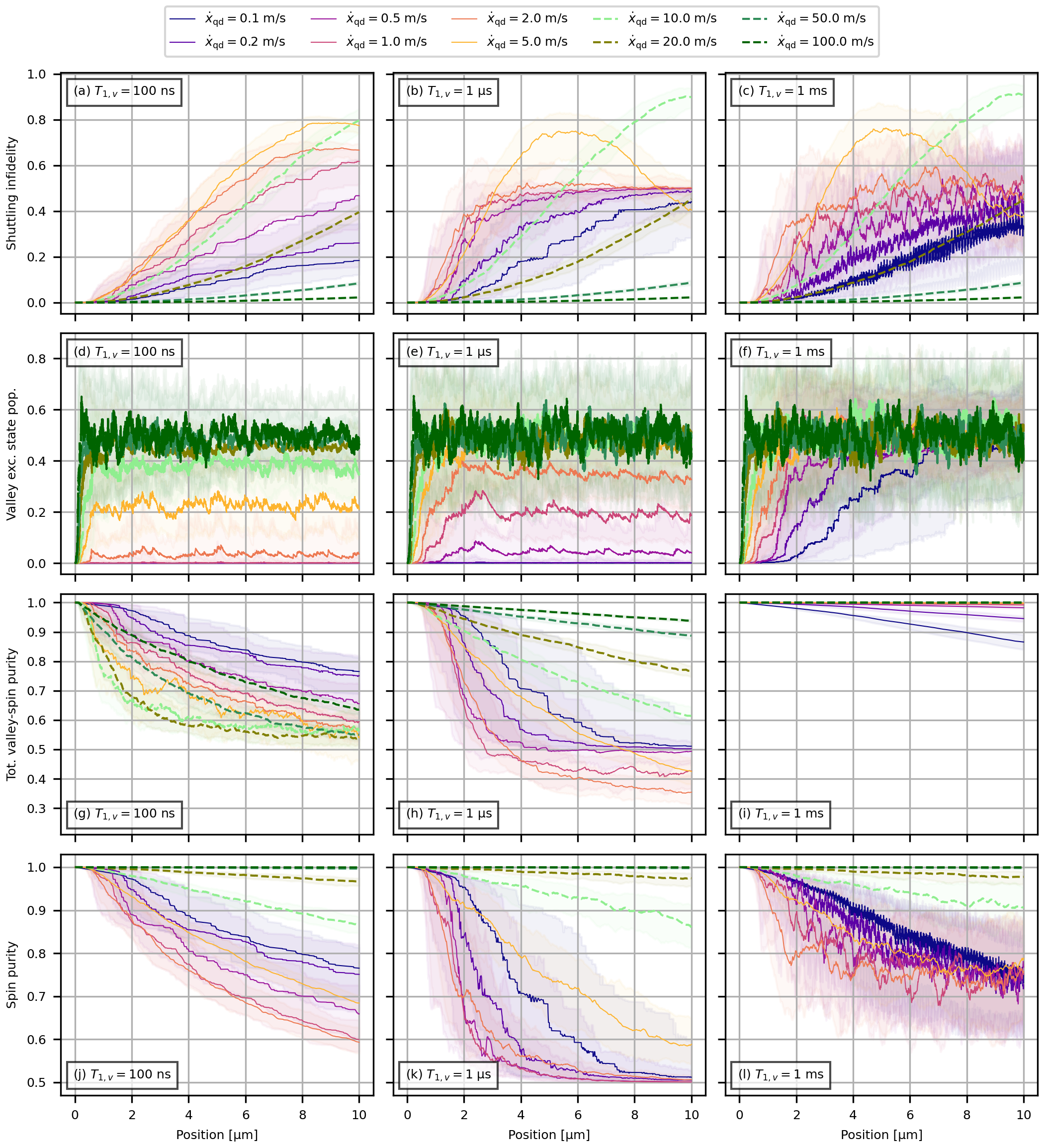}
    \caption{Spin transport infidelity (first row), population of the excited valley state (second row), purity of total spni-valley state (third row), purity of spin state (fourth row) for shuttling at different constant speeds as a function of the position along the shuttler. The effect of motional narrowing is included as an effective spin diffusion (coherence length 20 nm; coherence time 20 \textmu s). The different panels have valley lifetime (a) $T_{1,v} = 100$ ns, (b) $T_{1,v} = 1$ \textmu s, (c) $T_{1,v} = 1$ ms. Solid lines indicate the 50th percentile of 100 traces obtained from 100 different valley landscape realizations generated by the alloy diffusion model. Shaded area indicate the 25th to 75th percentile range for the corresponding color.}
    \label{fig:multi_plot_const_speed_complete}
\end{figure*}

In Fig.~\ref{fig:multi_plot_const_speed_complete} we lay out for each row from top to bottom the behavior of four key values during shuttling, namely, the spin shuttling infidelity $1-F_\text{ent}$ (see Eq.~\eqref{eq:shuttlingfidelity}), the population of the excited valley state $\bra{e(x_\text{qd})} \mathrm{tr}_S (\rho(t)) \ket{e(x_\text{qd})}$, the total purity $\mathrm{tr}(\rho(t)^2)$ and the purity of the spin state $\mathrm{tr}([\mathrm{tr}_V(\rho(t))]^2)$, where $\mathrm{tr}_S$ and $\mathrm{tr}_V$ are the partial traces over the spin and valley subspace respectively. 

We investigate three different valley lifetime scenarios, one for a short lifetime of $T_{1,v}$ = 100~ns (left column), one for an order of magnitude larger $T_{1,v}$ = 1~\textmu{}s (central column) and one for a long lifetime of $T_{1,v}$ = 1~ms (right column) close to experimentally measured values \cite{penthorn_direct_2020}.
These three regimes induce qualitative differences in the shuttling infidelity; in Fig.~\ref{fig:multi_plot_const_speed_complete}(a) we see that the infidelity increases monotonically in the range of speeds considered, even above 0.5; in Fig.~\ref{fig:multi_plot_const_speed_complete}(b) the rate of increment is higher and the infidelity saturates at 0.5 with the exception of $v$ = 5~m/s which shows a coherent oscillation; in Fig.~\ref{fig:multi_plot_const_speed_complete}(c) all speeds show coherent oscillations. The errors are consistently high for all parameters already at 10\% of shuttling distance.

The explanation comes from the high number of valley splitting minima associated with large valley phase difference. This induces diabatic transitions in the excited valley state with high probability, leading to different spin precession rates between the excited and ground state valley components and, therefore, valley-spin entanglement. Moreover, the interaction of the valley state with the phononic bath \cite{tahan_relaxation_2014} induces decoherence on both spin and valley states increasing the shuttling infidelity. In what follows we analyse the other key values in order to justify which mechanism is predominant for the shuttling infidelity.

Panels (d) and (e) indicate that the valley excited population has an average saturation value that depends on shuttling speed and valley relaxation time. This follows from the generally uniform distribution of valley splitting minima; crossing this minima induces excitations with increasing probability at higher speeds similarly to Landau-Zener transitions, but in between minima the valley has time to relax and the process reaches an equilibrium. This phenomenon could be actually used in experiments to determine the valley relaxation rate or the average distance between valley minima by measuring the average excited valley population at different speeds. From panel (f) we notice though that, as the valley lifetime increases, the valley population has time to build up reaching half filling which is the maximum average value given by the multiple random scatterings between ground and excited states. This maximum value is nevertheless reached with different rates (slopes) for different speeds. In panels (g) and (h) we show the loss of coherence in the joint valley-spin state induced by the valley relaxation dynamics. We notice that the loss is significant even for speeds that show valley population very close to zero. The case of $T_{1,v}$ = 1~\textmu{}s is intermediate as a longer relaxation time allows for a higher average valley excited population, as mentioned above, but at the same time it leads to a faster loss of coherence, which explains the higher rate of increment in panel (b) with respect to panel (a). In the case of $T_{1,v}$ = 1~ms, panel (i), the relaxation time is long enough that we can consider the total state to be pure, especially for speeds equal or higher than 1~m/s. Finally we can attribute the loss of spin purity in panels (j) and (k) predominantly to the valley relaxation, whereas in panel (l) the spin subsystem is entangled to the valley and the coherent oscillation saturate to an average spin purity value of 0.75.

For an entangled and pure valley-spin states,
\begin{equation}
    \mathrm{\rho_S (t)^2} = \frac{1}{2} \bigg ( 1 + \cos^2 \frac{2\kappa_z t}{\hbar} + (|\alpha|^2 - |\beta|^2)^2 \sin^2 \frac{2\kappa_z t}{\hbar} \bigg )
\end{equation}
$|\beta|^2 = 1 - |\alpha|^2$ is the population of the excited valley state.

The role of motional narrowing is important for the optimization described in Sec.~\ref{sec:optimization}. Here we follow the same derivation as in Ref.~\cite{langrock_blueprint_2023} and we translate the dephasing mechanism in an effective $T_{\phi,s}$ that we use in Eq.~\eqref{eq:master}. The results of Fig.~\ref{fig:multi_plot_const_speed_complete} include motional narrowing of coherence length $l_\text{coh}$ = 20~nm and static coherence time $T_2^*$ = 20~\textmu{}s. The conclusion of our analysis, presented in Appendix~\ref{sec:motionalnarrowing}, is that only speeds below 1~m/s are affected by an additional dephasing error. As our model does not include a microscopic or position dependent mechanism for this average decoherence, we focus on the optimization of speeds equal or higher than 1~m/s and from now on we consider the spin subspace to have infinite coherence.

\end{document}